\documentclass[prb,twocolumn,showpacs,preprintnumbers,amsmath,amssymb]{revtex4}
\usepackage{dcolumn}% Align table columns on decimal point
\usepackage{bm}% bold math
\usepackage{epsfig}
\begin{document}

%\wideabs{

\title{Direct demonstration of circulating currents in a controllable $\pi$-SQUID
generated by a 0 to $\pi$ transition of the weak links.}

\author{J.J.A.Baselmans, B.J. van Wees, and T.M. Klapwijk
$^\ast$} \affiliation{Department of Applied Physics and Materials
Science Center, University
of Groningen,\\ Nijenborg 4, 9747 AG Groningen, The Netherlands \\
$^\ast$Department of Applied Physics and DIMES, Delft University of
Technology, \\ Lorentzweg 1, 2628 CJ Delft, The Netherlands \\}
\date{\today}

\begin{abstract}
A controllable $\pi$-SQUID is a DC SQUID with two controllable
$\pi$-junctions as weak links. A controllable $\pi$-junction
consists of a superconducting - normal metal - superconducting
Josephson junction with two additional contacts to the normal
region of the junction. By applying a voltage $V_c$ over these
contacts it is possible to control the sate of the junction, i.e.
a conventional ($0$) state or a $\pi$-state, depending on the
magnitude of $V_c$. We demonstrate experimentally that, by putting
one junction into a $\pi$-state, a screening current is generated
around the SQUID loop at integer external flux. To be able to do
this, we have fabricated controllable $\pi$-junctions, based on
Cu-Nb or Ag-Nb, in a new geometry. We show that at 1.4 K only the
Nb-Ag device shows the transition to a $\pi$-state as a function
of $V_c$ consistent with theoretical predictions. In a
controllable $\pi$ SQUID based on Nb-Ag we observe, a part from a
screening current at integer external flux, a phase shift of $\pi$
of the $V_{SQUID}-B$ oscillations under suitable current bias,
depending on the magnitude of $V_c$.
\end{abstract}
\pacs{}%}
%introduction
%
\maketitle

\section{introduction}
In recent years considerable attention has been paid to
controllable Josephson junctions. In general, such a Josephson
junction consists of a Superconductor-normal metal-superconductor
(SNS) junction with additional electrodes connected to the normal
region of thew junction. Inside the normal region coherent quantum
states are formed due to the presence of the two superconducting
electrodes. The properties of the junction, and in particular the
critical current $I_c$ are determined by the energy spectrum
\emph{and} the occupation of these states. Hence it is possible to
change $I_c$ by changing the energy distribution of the
quasiparticles in the junction normal region
\cite{Volkov1,Volkov2,Wilhelm,Yip,BartIc}. This is done by sending
a control current through the additional electrodes connected to
the normal region or by applying a control voltage $V_c$ over
them. Several experiments have been performed the past years on
these devices, which differ mainly in the way in which the energy
of the quasiparticles is changed
\cite{Alberto,Schapers,Kutchinsky,me,Delsing}. The most
interesting situation is that of the so called controllable
$\pi$-junctions \cite{me,Delsing}. In these systems it is not only
possible to change the magnitude of the critical current of the
junction but also to reverse its direction with respect to the
phase difference $\varphi$ between the two superconducting
electrodes. This sign reversal is equivalent to the introduction
of an extra phase factor $\pi$ in the Josephson supercurrent
($I_{sc}$)-phase relation
\begin{equation}
   I_{sc}=I_c sin(\varphi) \Rightarrow I_{sc}=-I_c sin(\varphi)\equiv I_c sin(\varphi+\pi)
   \quad .
\label{Josephson}
\end{equation}
This means that for zero supercurrent, the ground state is that of
a macroscopic phase difference equal to $\pi$. There is an analogy
here with $\pi$-junctions associated with bi-crystals \cite{Tsuei}
or 's-d' contacts in ceramic superconductors \cite{Harlingen} and
$\pi$-junctions using a dilute ferromagnet f' as the "normal"
region of a S-f'-S junction \cite{Ryazanov,He3}. However, the
ceramic systems remain a $\pi$-junction (or normal junction) once
fabricated and the S-f'-S junctions switch between a normal and a
$\pi$-state only as a function of temperature. This in contrast
with the controllable $\pi$-junction where the state of the
junction as well as the magnitude of the critical current is fully
controllable by means of the control voltage.
\\
\\
Using such controllable $\pi$-junctions as weak links of a DC
SQUID (direct current superconducting quantum interference
device), it is possible to tune the critical currents of the
individual weak links. More interestingly, it is possible to add
an extra phase of $\pi$ or 2$\pi$ to the SQUID loop without the
application of an external magnetic field by switching one or two
of the junctions in a $\pi$ state. An extra phase factor of $\pi$
is identical to the external application of half a flux quantum,
$\phi_0 /2$, with $\phi_0 = 2.07 \cdot 10^{-7}$ Wb. This is a
direct consequence of the condition of a single-valued wave
function around the SQUID loop,
\begin{equation}\label{SQUIDeq}
\frac{2 \pi \phi}{\phi_0}-\varphi_1-\varphi_2= 2 \pi n \quad ,
\end{equation}
where $\phi$ is the total flux in the ring, n is an integer and
$\varphi_{1}$ and $\varphi_{2}$ are the phases over the two
Josephson junctions, as shown in Fig.\ref{SQUIDschema}a. Switching
one junction in the $\pi$-state at an external flux $\phi=n
\phi_0$ adds $\pi$ to the left hand side of Eq.\ref{SQUIDeq}. This
leads to the same solution of the equation if the junction would
be in the "conventional" state with an external flux $\phi=(n \pm
\frac{1}{2})\phi_0$. Hence it is possible to generate a screening
current in a SQUID with a suffici\"{e}ntly large inductance in the
absence of a magnetic field.
\\
\\
The main purpose of this paper is to demonstrate unequivocally
that it is possible to add a macroscopic quantum phase of $\pi$ or
2$\pi$ to a DC SQUID with 2 controllable $\pi$-junctions as weak
links. We demonstrate that this results in a screening current
that can be switched on or off depending on $V_c$ at an external
flux of an integer flux quantum.
\\
\\
The outline is the following: We begin, in section II, with a very
short description of the basics of the DC SQUID and a description
of the theoretical concepts regarding controllable
$\pi$-junctions. In section III we discuss the sample design and
fabrication and in section IV we discuss experiments on single
controllable $\pi$-junctions made of Nb-Ag or Nb-Cu. The geometry
of these devices differs from the conventional cross shape. We
show that the Nb-Ag device shows a 0 to $\pi$ transition at 1.4 K,
in contrast with the Nb-Cu device. The full dependence of the
critical currents of both devices on $V_c$ is in good quantitative
agreement with the theoretical predictions discussed in section
II. In section V we proceed with measurements on a controllable
SQUID, made of Ag-Nb. We show that the magnitude and sign of the
voltage oscillations of the SQUID as a function of the magnetic
field under suitable current bias can be tuned. In the last
section we measure directly the flux expelled by the screening
current in a controllable SQUID by placing a smaller SQUID
directly behind it. We show that the screening current around
integer external flux can be switched on or off by switching one
of the junctions of the controllable SQUID in a $\pi$- or a normal
state.

\section{theoretical concepts}

\subsection{the DC SQUID}

\begin{figure}[t]
\centerline{\psfig{figure=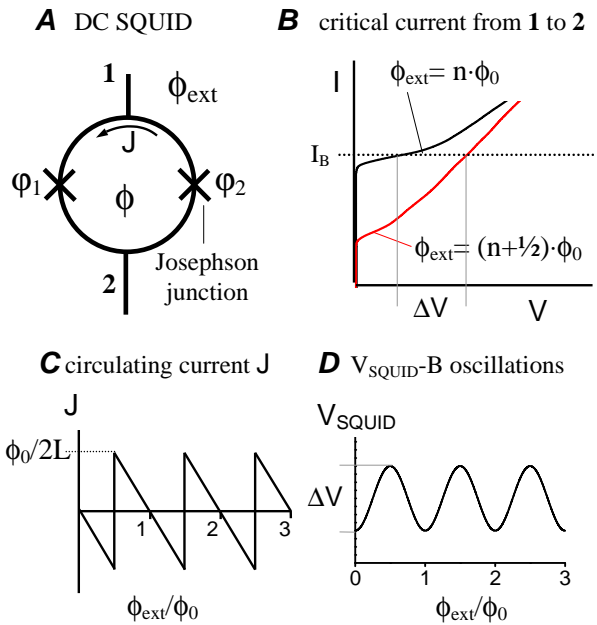,width=8 cm,clip=}}
\caption{Basic operation of the DC SQUID. (A) shows a DC SUID with
a screening current J due to an external flux $\phi_{ext}$. The
supercurrent of the SQUID, as shown in (B), oscillates as a
function of $\phi_{ext}$. By applying a current bias $I_B$ the
oscillating critical current is transferred into an oscillating
voltage $V_{SQUID}$ between contacts 1 and 2 (A), as shown in (D).
The screening current J has, in the limit of large self inductance
L, a saw tooth shape as a function of $\phi_0$, as shown in (C). }
\label{SQUIDschema}
\end{figure}

Before we discuss the controllable $\pi$-junction it may be
helpful to discuss the basics of the DC
SQUID\cite{Clarke,Pavuna,Tinkam}. A DC SQUID is a ring of
superconducting material with self inductance L and two Josephson
junctions with a critical currents $I_{c1}$ and $I_{c2}$. In the
absence of any external flux, $\phi_{ext}$, the SQUID is in
equilibrium and behaves as a single Josephson junction with a
total critical current $I_c$= $I_{c1}$+$I_{c2}$. If $\phi_{ext}
\neq 0$, then the total flux $\phi$ in the SQUID loop changes,
depending on its self inductance, according to
\begin{equation}
\phi = \phi_{ext}+LJ \label{fluxeq} \quad ,
\end{equation}
with $J$ the screening current running around the SQUID loop. The
flux in the loop, $\phi$, also depends on the phases of the two
Josephson junctions
\begin{equation}
\varphi_1 = \frac{2 \pi \phi}{\phi_0} -\varphi_2 \label{phaseeq}
\quad ,
\end{equation}
which is directly obtained from Eq.\ref{SQUIDeq}. In the limit of
small self inductance, so that the $LJ$ term in Eq.\ref{fluxeq} is
negligible and if $I_{c1}=I_{c1}\equiv I_{c1,2}$, it is easy to
show that
\begin{equation}
I_c=2I_{c1,2}\left|cos\left(\frac{2 \pi \phi_{ext}
}{\phi_0}\right)\right| \quad ,\label{1}
\end{equation}
where we use Eqs.\ref{Josephson},\ref{fluxeq} and \ref{phaseeq}.
In this case $I_c$ is periodic in $\phi_0$ and varies between
$2I_{c1,2}$ at $\phi_{ext}=n \phi_0$ and 0 at $\phi_{ext}=(n+1/2)
\phi_0$. At nonzero inductance the periodicity remains the same
but the amplitude of the $I_c$ modulation is reduced, for $I_c$=0
is not reached at $\phi=\pi/2$ due to the presence of the
circulating current $J$. In the limit of large self inductance,
the amplitude of the $I_c$ modulation decreases to $\phi_0/L$,
whereas $J$ becomes a saw tooth function varying linearly from $-
\phi_0/2L$ to $+ \phi_0/2L$, as shown in Fig.\ref{SQUIDschema}(B)
and (C). The periodic behavior of $I_c$ as function of
$\phi_{ext}$ can be transformed into a periodic voltage by sending
a bias current $I_B$ a little larger than the maximum value of
$I_c$ through the SQUID. Due to the non-linear I-V curve of the
SQUID the difference in $I_c(\phi_{ext})$ translates into a
voltage $V_{SQUID}(\phi_{ext})$, thus creating a very sensitive
flux to voltage transformer. The resulting $V_{SQUID}-B$
oscillations are shown in Fig.\ref{SQUIDschema}(D).

\subsection{Controllable $\pi$-junctions}

\begin{figure}[t]
\centerline{\psfig{figure=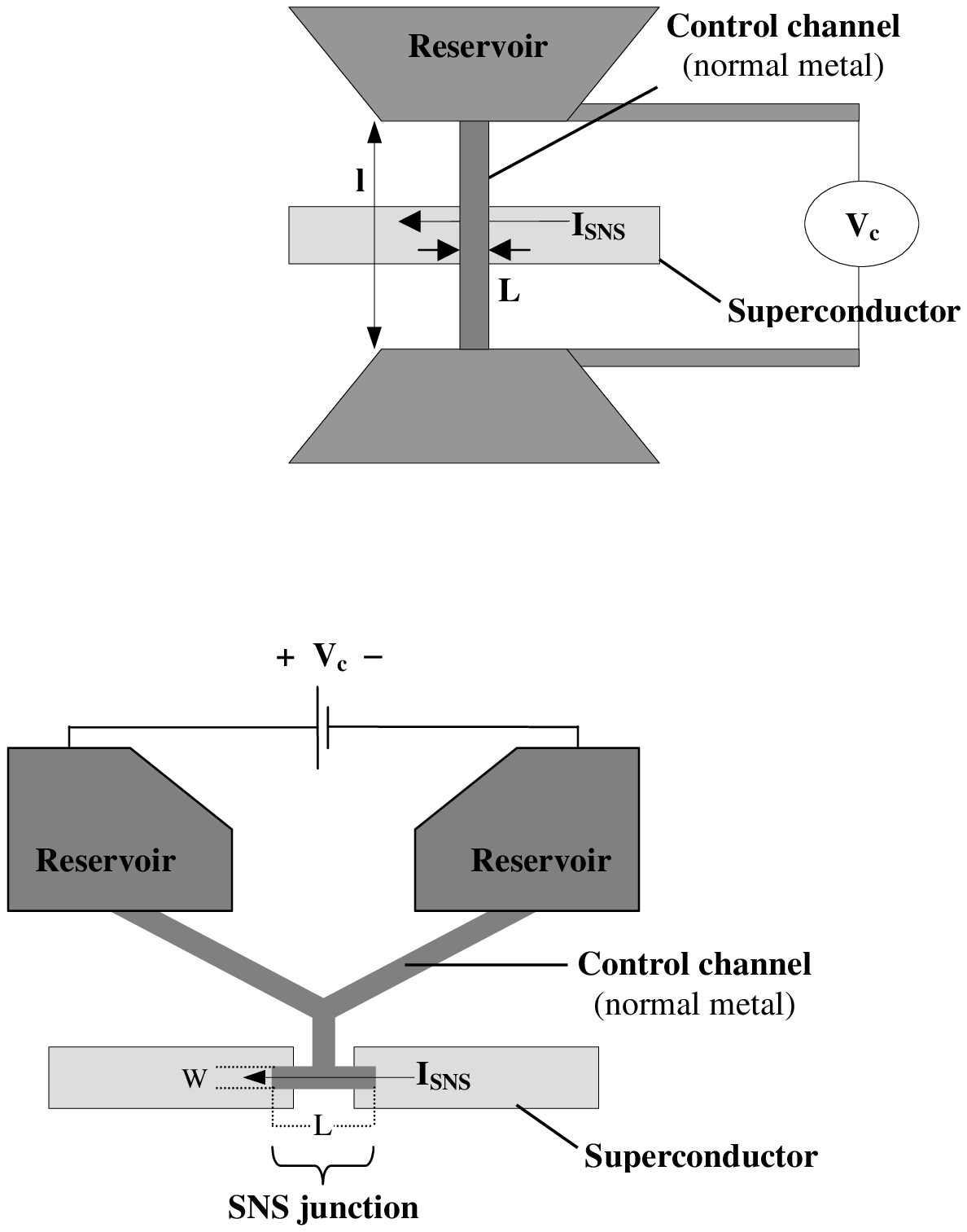,width=8 cm,clip=}}
\caption{Schematic drawing of a controllable $\pi$-junction. The
center of the control channel, which connects two large
reservoirs, is coupled to the normal region of a co-planar SNS
junction. By applying a voltage $V_c$ over the reservoirs the
energy distribution of the quasiparticles in the channel and hence
in the normal region of the SNS junction is changed. As a result
the occupation of the suercurrent carrying states in N is changed,
resulting in a change in the critical value of the supercurrent
$I_{SNS}$.} \label{schema}
\end{figure}

A controllable $\pi$-junctions is a device that combines two
different structures in one: A thin film SNS junction and a short
mesoscopic wire (which we call control channel) between two large
electron reservoirs in thermal equilibrium. A schematic drawing is
given in Fig.\ref{schema}: The center of the control channel,
which connects two large reservoirs, is coupled to the normal
region of a co-planar thin film SNS junction, in which the normal
region partly overlaps the superconducting electrodes. The length
and width of the normal region are denoted by L and W in the
figure. The size of the junction $L \lesssim 1\mu m$ and the
control channel length, $l \sim 5\mu m$, is larger than the
elastic mean free path of the electrons, which is approximately
given by the film thickness ($\sim 50nm$). Hence the electron
motion is diffusive.
\\
\\
The principle of operation is the following: The presence of the
two superconducting electrodes induces a continuous spectrum of
supercurrent carrying states $Im(J(E))$ in the normal region of
the junction, which is responsible for the supercurrent transport
($I_{SNS}$ in Fig.\ref{schema}). The energy scale $E$ is the
energy with respect to the Fermi level. As stated before, the
critical value of $I_{SNS}$, $I_c$, is determined by the energy
spectrum \emph{and} the occupation of these supercurrent carrying
states. The occupation of the states is determined by the energy
distribution of the quasiparticles in the normal region of the
junction, $f(E)$, which is identical to the energy distribution in
the center of the control channel. This energy distribution can be
modified by means of applying a voltage $V_c$ over the control
channel, so $f(E)=f(E,V_c)$. The relation between $I_c$, $Im(J(
E))$ and $f(E,V_c)$ is given by
\cite{Volkov1,Volkov2,Wilhelm,Yip,BartIc}:
\begin{equation}
I_{c}={\frac{1}{R_{n}}\int_{-\infty}^{\infty}}dE[1-2f(E,V_c)]Im(J(E,\varphi=\pi/2))
\medspace, \label{IcRn}
\end{equation}
where $\varphi$ the phase difference between the superconducting
electrodes \cite{critcur}. It is important to realize that the
critical current is modified by changing the energy of the
quasiparticles in the normal region and not by the flow of the
control current associated with $V_c$. In the present geometry
this is quite clear, because the current paths of $I_{SNS}$ and
the control current are separated. This is in contrast with the
conventional cross geometry, where $I_{SNS}$ crosses the control
current. However, even in this case it is the energy of the
quasiparticles and not the control current that determines the
critical current.

\begin{figure}[t]
\centerline{\psfig{figure=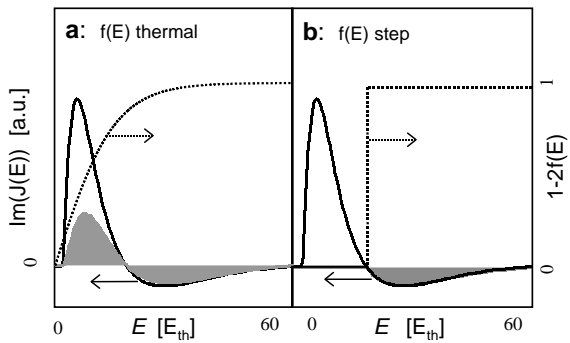,width=8 cm,clip=}}
\caption{Supercurrent carrying density of states $Im(J(E))$ for
$\Delta$/$E_{th}$=70 at $\varphi$ = $\pi$/2 (solid line) together
with $1-2f(E)$ (dotted line) in the case of a thermal distribution
(a), and a step distribution (b). The shaded area represents the
contributions to the integrand of Eq.\ref{IcRn}. $E$ represents
the energy with respect to the Fermi level an $E_{th}$ is the
Thouless energy.} \label{scdos}
\end{figure}

To calculate the dependence of $I_c$ as a function of $V_c$ we
need to know both the shape of the supercurrent carrying density
of states, $Im(J(E))$ and the shape of the quasiparticle energy
distribution in the normal region of the junction, $f(E, V_c)$.
The exact energy dependence of $Im(J(E))$ is calculated using the
quasi-classical Green's function theory
\cite{Volkov1,Volkov2,Wilhelm,Yip}. It depends on the ratio of the
Thouless energy with the energy gap $\Delta$ of the
superconductors. The Thouless energy is given by $E_{th}=\hbar
D/L^2$, with L the length of the normal region of the SNS
junction\cite{Dubos} and D the diffusion constant of the normal
metal. The general shape of $Im(J(E))$, as shown by the solid
black line in both panels of Fig. \ref{scdos} is a strongly damped
oscillation with a hard gap at low energies $E\lesssim E_{th}$.
The positive and negative parts of the supercurrent carrying
density of states represent energy dependent contributions to the
supercurrent in the positive and negative direction.

The other relevant quantity of the system is the quasiparticle
energy distribution, $f(E,V_c)$. The exact shape of $f(E,V_c)$ in
the center of the control channel (and hence in the junction
normal region) depends, as shown by Pothier \textit{et al.}
\cite{Pothier}, on the voltage $V_c$ applied over the control
channel and on the amount of electron-electron and electron-phonon
scattering. At low temperatures, $T\lesssim 1K$, electron-electron
scattering is the dominant relaxation mechanism and the electron
energy distribution depends only on $V_c$ and the
electron-electron interaction. In this case, we can distinguish
two limiting possibilities:
\\
\\
\textit{1: no electron-electron interaction} If the control
channel is sufficiently short, so that the electron-electron
interaction time $\tau_0$ exceeds the diffusion time $\tau_D$
through the channel, the electrons conserve their energy. The
resulting electron energy distribution in the center of the wire
is given by\cite{Pothier}
\begin{equation}
f(E, V_c)=1/2 \cdot \left[f_d\left(E-\frac{e
V_c}{2}\right)+f_d\left(E+\frac{e V_c}{2}\right)\right] \quad ,
\label{step}
\end{equation}
here $f_d(E)$ is the Fermi-Dirac distribution at a temperature T
\begin{equation}
f_d(E)=\left[1+exp\left(\frac{E}{k_B T}\right)\right]^{-1} \quad
,\label{fd}
\end{equation}
with $k_B$ Boltzmann's constant. Eq.\ref{step} is the renormalized
addition of the thermal distribution functions of the reservoirs
with an energy separation of $eV_c$. This represents, if
$k_bT<eV_c$, a double step structure with an occupation fraction
of $1/2$ in an energy range of $eV_c$ around the Fermi energy.
Such a step distribution will, at a large enough value of $V_c$,
block all the positive contributions of $Im(J(E))$, as shown in
Fig. \ref{scdos}b, because an occupation fraction of $1/2$ causes
the integrand of Eq.\ref{IcRn} to vanish. This results in a
reversal in the direction of the
supercurrent and hence the transition to a $\pi$-state\\
\\
\textit{2: very strong electron-electron interaction} If the
control channel is very long, so that $\tau_0 \ll \tau_D$, the
electron system regains a \emph{local} thermal equilibrium. The
energy distribution given by the Fermi-Dirac distribution, with an
effective temperature $T_{eff}>T_{Bath}$,
\begin{equation}
T_{eff}=\sqrt{{T_{bath}}^2+\left(aV_c\right)^2} \quad
.\label{Teff}
\end{equation}
Here $T_{bath}$ is the bath temperature and $a$ can be calculated
using the Wiedemann-Franz law, resulting in $a$=3.2 K/mV. This
situation is called the 'hot electron limit'. The result, a
Fermi-Dirac distribution with a width increasing with $V_c$,
causes a rapid decay of the supercurrent, as shown in the left
panel of Fig.\ref{scdos}a due to the compensation of the positive
and negative contributions in $Im(J(E))$.
\\
\\
The shape of the distribution function in a real experiment is in
general in between these two limits, with possible modifications
due to electron-phonon interactions. We calculate the exact shape
of the energy distribution function using a model presented by
Pothier \textit{et al.}\cite{Pothier,Pierre}. In this model the
shape of the electron energy distribution in the wire depends on
four parameters: The voltage $V_c$ applied over the wire, the
effective strength of the electron - electron interaction, the
effective strength of the electron phonon interaction and possible
heating of the reservoirs. The electron-electron interaction is
described by a two-particle interaction with an interaction Kernel
$K_{\alpha} \epsilon^{-\alpha}\cdot\tau_D$. Here $\epsilon$ is the
energy transferred between two interacting particles and $\tau_D$
the diffusion time through the wire. The magnitude $K_{\alpha}$
and energy dependence $\alpha$ of the electron - electron
interaction can be calculated by means of a direct calculation of
the screened Coulomb interaction in a homogenous 1-dimensional
diffusive conductor \cite{Altshuler1,Altshuler2}. Without going
into detail it is important to realize that a direct measurement
of the electron energy distribution function in diffusive 1D gold,
copper and silver wires using tunnel junction spectroscopy
\cite{Pothier,Pierre,Gougam,Pierregold} yields material dependent
results for $K_{\alpha} \epsilon^{-\alpha}$ which are not
consistent with the theoretical predictions using
refs.\cite{Altshuler1,Altshuler2} alone. Because the control
channels used in our experiment are similar to the wires measured
in refs.\cite{Pothier,Pierre,Gougam,Pierregold}, we have used the
experimental values of $\alpha$ in our calculations: $\alpha=2$
for copper and $\alpha=1.2$ for Ag. The strength of the
interaction, $K_{\alpha}$ is used as a fit parameter.

In the model, the effect of electron-phonon scattering is
implemented in a similar way. The magnitude of the interaction
kernel is, in this case, deduced from measurements of the phase
relaxation time\cite{Pierrethesis}, and given by
$K_{phonon}=8ns^{-1}meV^{-3}$ for Cu and Ag.

As a last input parameter, possible heating of the reservoirs
should be taken into account. The power associated with the
application of the control voltage, $P=V_c^2/R_c$ ($R_c$ is the
control channel resistance) is injected into the reservoirs. At
low bath temperatures (T $<$ 1K), low values of $R_c$ and large
values of $V_c\gtrsim 1mV$, the electron gas in the reservoirs can
attain a much higher effective temperature than the bath
temperature\cite{UrbinaClarke,Henny}. However, all experiments
presented in the remainder of the text are performed at relatively
high temperatures of 1.4 K, resulting in negligible heating of the
reservoirs \cite{commheating}.

\section{Sample design and fabrication}

\begin{figure}[t]
\centerline{\psfig{figure=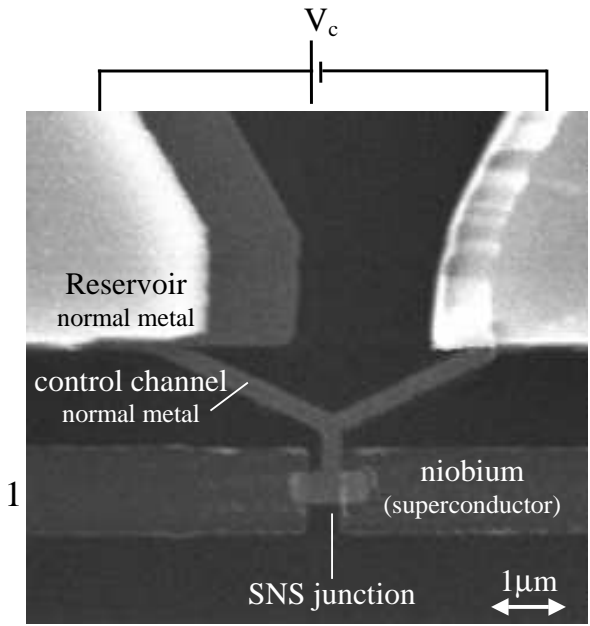,width=8 cm,clip=}}
\caption{Scanning electron micrograph of a controllable
$\pi$-junction using side contacts. In this case the normal metal
is silver, whereas also similar samples using copper are made.}
\label{jie}
\end{figure}

So far, all experimental evidence of a controllable $\pi$-junction
\cite{me,Delsing} has been achieved using devices with a cross
geometry. The control channel and the normal region of the
junction are in fact a single cross shaped metallic wire, with
superconducting contacts at the first two opposite ends and large
reservoirs at the others. This geometry is not suitable for use in
a SQUID geometry, because one of the reservoirs of each junction
would have to be within the SQUID loop, which is not possible due
to the large reservoir sizes needed. Therefore we use a device
geometry with side contacts (see Fig.\ref{jie}), such as used by
Morpurgo \emph{et al.}\cite{Alberto} in the thermal limit. The
disadvantage of this geometry is that it imposes a larger minimum
length on the control channel than the conventional cross
geometry, where this length can be reduced to about 1 $\mu$m. As a
consequence, a material with a slow electron-electron relaxation
is needed to be able to maintain a non-thermal energy distribution
in the control channel. For this reason gold is unsuitable as the
normal metal\cite{Pothier,Pierre,Gougam,Pierregold,menew} and
copper or preferably silver should be used.

We now describe the sample fabrication, referring to Fig.
\ref{jie}. The samples have been realized on a thermally oxidized
Si wafer that is covered with a 150 nm layer of sputter deposited
$Al_2O_3$ to improve the adhesion of Ag and Cu. In the first step
50 nm of Nb is deposited using standard e-beam lithography, DC
sputtering and subsequent lift-off. The critical temperature of
the sputtered film is 8.1 K. Subsequently the normal region,
control channel and the reservoirs (Ag or Cu) are deposited using
shadow evaporation without breaking the vacuum. This is needed
because the adhesion of these metals is poor, implying that it is
not possible to bake a deposited film to be able to do another
lithography step. We use a double layer of PMMA-MA and PMMA with
e-beam lithography and wet etching to create a PMMA suspended
mask. The deposition of the metals is done in an UHV deposition
system with a background pressure of $5\cdot 10^{-10}$ mBar, the
pressure in the system during the evaporation steps is $\leq
5\cdot 10^{-8}$ mBar. Prior to deposition we use Argon etching
($P_{Ar}=1\cdot 10^{-4}$ mBar, 500 V) for 3.5 minutes to clean the
Nb surface. Subsequently we deposit, in the case of an Ag sample,
10 nm of Ti adhesion layer under a large angle ($47^{\circ}$,
resulting in a shift to the left in the figure), with the result
that the Ti layer is only deposited on the substrate at the
position of the reservoirs and not at the position of the thin
openings defining the control channel and the normal region of the
junction. This is important, for the effect of a thin Ti layer
upon the electron-electron interaction time is not known. Then we
deposit 50 nm Ag perpendicular to the substrate, thus creating the
control channel and the normal region of the junction. As a last
step 700 nm of Ag is deposited again at $47^{\circ}$ to form the
reservoirs with an effective thickness of 475 nm. In Fig.\ref{jie}
the left shift due to the large evaporation angle of the
reservoirs with respect to the control channel is clearly visible.
A (parasitic) projection of the control channel and the normal
region of the SNS junction is not present, because the aspect
ratio of the top resist is so large for these structures that the
metal is deposited into the side wall of the resist. This results
in a removal of these projections during lift-off. The fabrication
of the Cu samples is identical except for the fact that Cu is used
for the normal conducting regions of the device and that no Ti
adhesion layer is used underneath the reservoirs. We found that
this is not necessary for sufficient adhesion. The length of the
control channel cannot be made shorter, because the resist will
become unstable if the large openings defining the reservoirs come
any closer \cite{commGe}. To measure the quality of the niobium -
normal metal interface we have made, in every fabrication run, a
cross of a 200 nm wide Nb and normal metal wire. The resistance of
the 200x200 nm interface has been determined to be $\approx$ 0.1
$\Omega$ each time, which is smaller than the square resistance of
the normal metal (0.4 $\Omega$), indicating that the interface is
clean.

\section{experiments on single junctions}

\begin{figure}[t]
\centerline{\psfig{figure=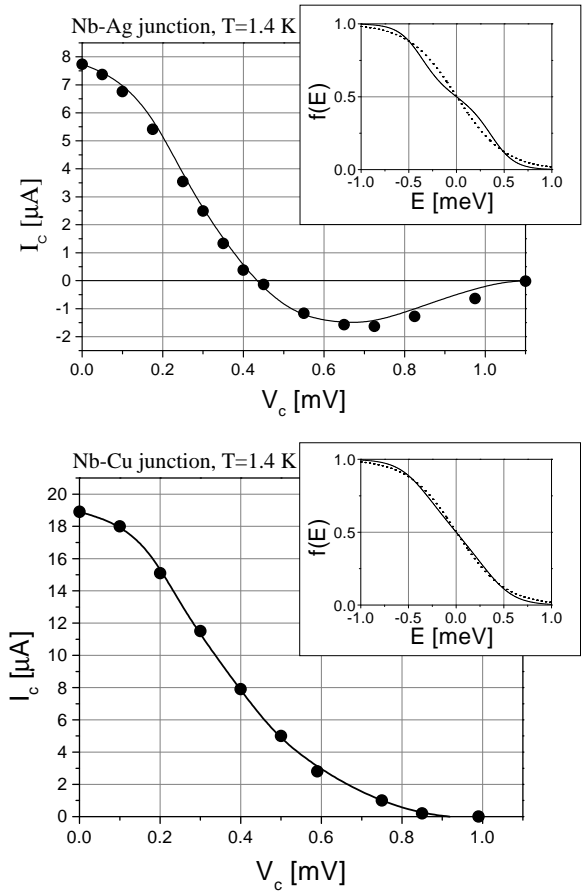,width=8 cm,clip=}}
\caption{Critical current of a controllable $\pi$-junction as
shown in Fig.\ref{jie}, made from Cu-Nb (bottom) and Ag-Nb (top).
The dots represent the data and the solid line the calculated form
using Eq. \ref{IcRn}. The solid lines in the inset show the
electron distribution functions at $V_c$=0.8 mV used in the fit.
For comparison, the Fermi-Dirac distribution in the hot electron
regime, with $T_{eff}=2.9 K$, is given by the dotted line.}
\label{resjie}
\end{figure}

\begin{table}
\caption{\label{tab:tablejies} Parameters of the single SNS
junctions. $R_n$ is the normal state resistance of the SNS
junction, L is the length of the normal region and W the width
(see Fig.\ref{schema}). $E_{th}$ is the Thouless energy obtained
using D and L.}
\begin{ruledtabular}
\begin{tabular}{lllll}
{\bf sample} & L x W [nm] &  $R_{n}$ [$\Omega$] & D [m$^2$/s] &
$E_{th}$ [$\mu eV$]\\ \hline Ag & 1000x400 & 0.65& 0.023& 14
\\ Cu & 800 x 600 & 0.8 & 0.021 & 21\\
\end{tabular}
\end{ruledtabular}
\end{table}
In the experiment, performed at 1.4 K in a pumped He bath, we
measure the critical current of the SNS junction (from contacts 1
to 2) as a function of the voltage $V_c$ applied over the control
channel (see Fig. \ref{jie}). The measurement of the critical
current consists of sweeping the bias current while measuring the
differential resistance $dV/dI$ using a small AC modulation on the
bias current. We define the experimental critical current where
the differential resistance reaches half its normal state value
\cite{Dubos}. The control voltage is applied using another current
source, with floating ground and $V_c$ is measured by means of
voltage probes close to the control channel ends (not visible in
the figure). The measurement of $V_c$ shows a slight modulation if
the current bias through the SNS junction is smaller than the
critical current, associated with Andreev interferometry
\cite{Esteve,Bart}. This modulation is used to determine the state
of the junction\cite{me}. The experiment is performed on several
devices made of Cu-Nb and Ag-Nb. All samples show identical
behavior and we present the data of two samples, one made of Ag-Nb
and one made of Cu-Nb. The exact parameters of these two samples
are given in Table \ref{tab:tablejies}. The experimental critical
current as a function of $V_c$ is shown by the solid circles in
Fig. \ref{resjie}. From the figure is is obvious that the Ag-Nb
device shows a transition to a $\pi$-junction above
$V_{c}=V_{c,critical}=430 \mu V$ and that the device made of Cu-Nb
does not show this transition. The magnitude of the equilibrium
critical current at 1.4 K is, in the case of Ag, in good agreement
theoretical predictions presented by Dubos \textit{et
al.}\cite{Dubos}. The measured fraction of $I_cR_n/E_{th}=0.44$,
whereas 0.5 is predicted. For the copper sample the agreement is
less good. In this case we measure 0.5 whereas 2 is predicted. The
next step in our analysis is a quantitative comparison of the
experimental data with the theory. As input parameters to
calculate the distribution function, we use, for $\alpha$ and
$K_{phonons}$ the values given in section II\cite{Pothier,Pierre},
$K_{phonons}$=8 $ns^{-1}meV^{-3}$, $\alpha$=1.2 for Ag and 2 for
Cu. As stated before, reservoir heating is negligible due to the
relatively high bath temperature and large control channel
resistances ($\sim$8 $\Omega$ for both samples), despite the large
values of $V_c$, \cite{commheating}. Using $E_{th}$ and
$K_{alpha}$ as fit parameters, we obtain the best fit, as shown by
the black lines in both panels of Fig.\ref{resjie}, using the
parameters shown in Table \ref{tab:tableresult}. In the Table, we
have calculated $\tau_D=\frac{l^2}{D}$ using for l the total
diffusion length, given by the length of the v-shaped wire plus
two times the T-shaped extension towards the junction, yielding
l=5+2$\cdot$1=7$\mu m$ (see Fig.\ref{jie}.
\begin{table}
\caption{\label{tab:tableresult} Result of the fits on the single
junctions. For $\alpha$ we use the experimental values from
Ref.\cite{Pothier,Pierre}: $\alpha$=2 for Cu and 1.2 for silver.}
\begin{ruledtabular}
\begin{tabular}{llll}
{\bf sample} &{\bf $K_{\alpha}$ [$ns^{-1}\mu eV^{\alpha-2}$]}
&{\bf $E_{th}$ [$\mu eV$]} & {\bf $\tau_D$ [$ns$]}
\\ \hline Ag & 0.65 & 13 & 2.3
\\ Cu & 1.7 & 21.5 & 2.5 \\
\end{tabular}
\end{ruledtabular}
\end{table}
The fits are very good over the entire energy range. Moreover, the
results on $K_{\alpha}$ are in good agreement with the values
obtained in the experiments from \cite{Pothier,Pierre}, in
contrast to a similar analysis on a conventional $\pi$-junction
using gold \cite{menew}. The values of the Thouless energy
inferred from the fits are both in good agreement with the
geometrical values. The solid lines in the inserts of the figure
show the electron distributions, inferred from the fits, at
$V_c$=0.8 mV. For both samples the distributions are relatively
rounded. In the case of the Ag device, this is mainly caused by
the bath temperature of 1.4 K. In the case of the Cu sample the
stronger electron-electron interaction is responsible for the
extra rounding. However, the distribution is, even in this case,
not quite a thermal one but still has a weak double step
signature. This can be seen by the difference between the solid
and the dotted line, which represents a thermal distribution with
$T_{eff}=2.9 K$. This temperature corresponds to the effective
temperature in the hot electron regime, using Eq.\ref{Teff} with
$V_c$=0.8 mV and $T_{bath}$=1.4K. Model calculations indicate
that, in the case of the Cu-Nb device, a slight decrease in
control channel length or measuring at a lower bath temperature
would result in a distribution that is sufficiently less rounded.
This would result in a transition to a $\pi$-state at large enough
values of $V_c$\cite{commcu}.

We conclude that, in the present geometry, only a Nb-Ag junction
shows the transition to a $\pi$-junction at 1.4 K. The absolute
value of the critical current and the response of the critical
current to $V_c$ are all consistent with the experimental and
theoretical data available. A Nb-Cu junction does not show a
transition to a $\pi$-junction at this temperature, but would
probably do so at lower temperatures. To be able to measure at 1.4
K and higher temperatures we have fabricated controllable
$\pi$-SQUIDS, DC SQUIDS with controllable $\pi$- junctions as the
weak links, from Nb-Ag.

\section{controllable $\pi$-SQUID}
\begin{figure}[t]
\centerline{\psfig{figure=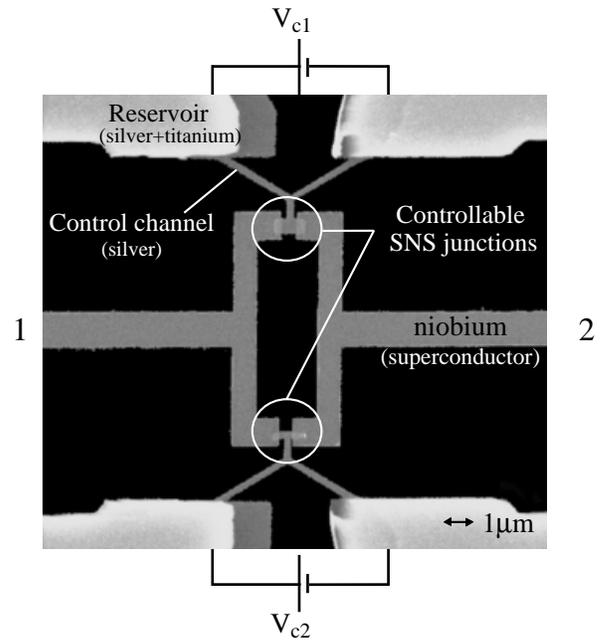,width=8 cm,clip=}}
\caption{Scanning electron micrograph picture of a controllable
$\pi$-SQUID.} \label{SQUID}
\end{figure}

A practical realization of a controllable $\pi$-SQUID made of
Nb-Ag is shown in Fig. \ref{SQUID}. The fabrication procedure is
identical to the fabrication of the single junctions described in
Section III. The length L of the normal regions of both junctions
is 1100 nm and the width W of the normal regions is 520 nm for the
top junction and 220 nm for the bottom junction. The resistance of
the control channels is 8 $\Omega$, yielding a diffusion constant
D=0.023 $m^2 /s$. From L and D, $E_{th}$ is estimated to be 12.6
$\mu eV$ for both junctions. The area of the SQUID loop is 12$\mu
m^2$. The Thouless energy of both junctions of the SQUID are
almost identical to the Thouless energy of the single junction
made of silver discussed in the previous paragraph. We therefore
conclude that the response of the critical current of each
junction to $V_c$ is comparable to the one presented in the top
panel of Fig. \ref{resjie}.

\subsection{Experiments on a controllable SQUID}
\begin{figure}[t]
\centerline{\psfig{figure=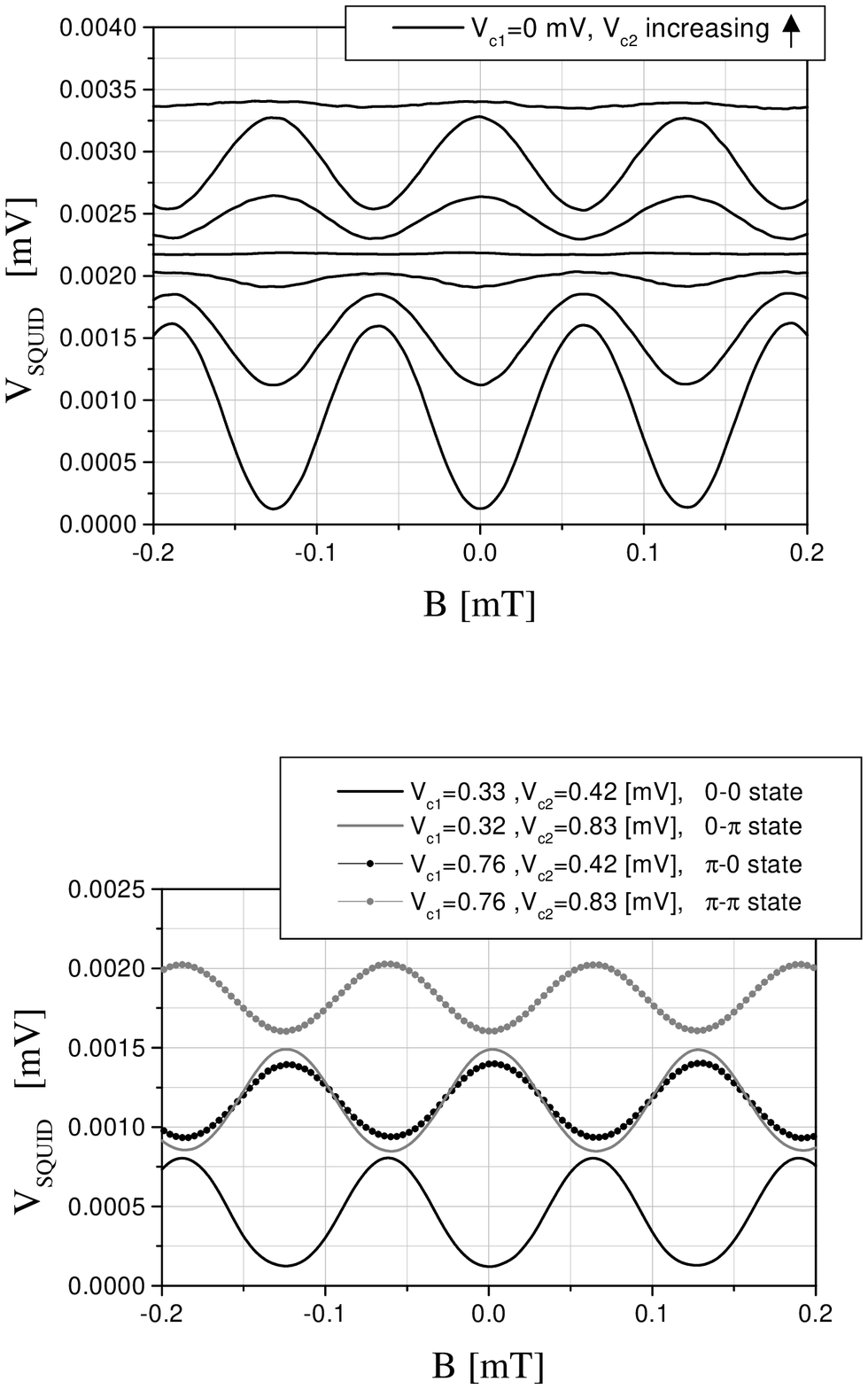,width=8 cm,clip=}}
\caption{The voltage over the controllable SQUID as a function of
the external magnetic field B using a current bias $I_B=4\mu A$
(curves offset for clarity). Solid lines: $V_{c1}$=0,
$V_{c2}$=0.30,0.38,0.46,0.48,0.54,0.70,2.7 mV (bottom to top).}
\label{oscillations}
\end{figure}

In the first experiment we measure the $V_{SQUID}-B$ oscillations
of the controllable SQUID at 1.4 K. We bias the SQUID with a low
frequency AC bias current ($I_{Bias}$,$~$80Hz, 4$\mu A$) between
contacts 1-2. The amplitude of the bias current is a little larger
than the maximum critical current of the SQUID. The voltage over
the SQUID is measured, $V_{SQUID}$, from contacts 1 to 2, as a
function of the applied magnetic field B using a lock-in
amplifier. This lock-in technique strongly reduces the noise
compared to the conventional measurement method using a DC bias.
Simultaneously we send a DC current through the bottom control
channel and measure the resulting control voltage $V_{c2}$, using
current sources with floating grounds. The result, using the
device shown in Fig \ref{SQUID}, is shown in Fig
\ref{oscillations}. The solid lines represents the $V_{SQUID}$-B
oscillations for increasing values of $V_{c2}$ ($V_{c1}=0$). At
first the amplitude of the oscillations decreases with increasing
$V_{c2}$ and reaches zero at $V_{c2,critical}$=0.48 mV, indicating
that the critical current of the bottom junction is equal to 0. At
higher values of $V_{c,2}$ the $V_{SQUID}$-B oscillations
re-appear, with a shift $\pi$ in phase with respect to the
oscillations at lower values of $V_{c2}$. The bottom junction and
hence the SQUID, are now in the $\pi$ state. At zero field we now
measure a voltage maximum in stead of a minimum.

\begin{figure}[t]
\centerline{\psfig{figure=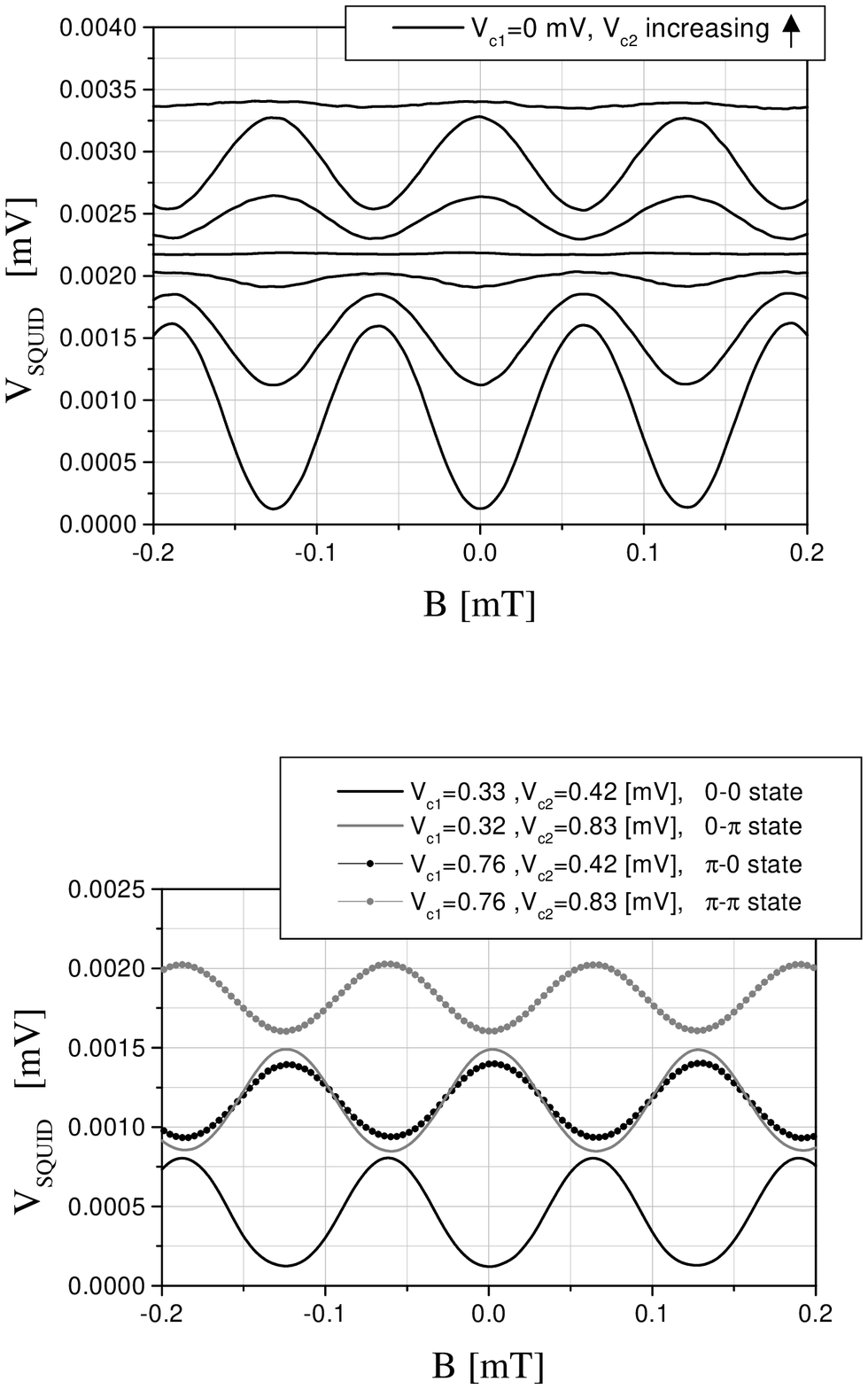,width=8 cm,clip=}}
\caption{The voltage over the controllable SQUID as a function of
the external magnetic field B (curves offset for clarity) using
both controllable junctions. The solid black line represents a 0-0
state, the solid grey line and the dotted black line represent the
situation with only junction 2 or 1 in the $\pi$-state,
respectively. The dotted grey line gives the situations for both
junctions in a $\pi$ state. Note that it has the same phase as the
solid black line.} \label{oscillations2}
\end{figure}

Using an additional current source it is possible to apply a
control voltage also to the top junction. Now it is possible to
add a phase of 0, $\pi$ or 2$\pi$ to the SQUID loop, as shown in
Fig.\ref{oscillations2}. The solid black line represent the 0-0
state, where both $V_{c1}$ and $V_{c2}$ are smaller than
$V_{c,critical}$. By putting junction 1 (dotted black line in
Fig.\ref{oscillations2}) \emph{or} junction 2 (solid grey line in
the figure) in a $\pi$-state by applying a sufficiently large
value of $V_c$ we observe in both cases an identical shift of
$\pi$ in phase. By putting \emph{both} junctions in a $\pi$ state,
as shown by the dotted grey line in Fig.\ref{oscillations2}, we
regain the original phase of the oscillations due to the 2$\pi$
periodicity of the macroscopic wave function around the SQUID
loop. Similar measurements at 100 mK in a dilution refrigerator
and at 4.2 K have shown similar results\cite{aplsquid}, with
however different amplitudes of the V-B oscillations due to the
temperature dependence of the critical current of the Josephson
junctions. These results indicate that behavior of the SQUID in
the $\pi$-state at $\phi_{ext}=n \phi_0$ is identical to that of a
conventional SQUID at $\phi_{ext}=(n+1/2) \phi_0$. This should
also be valid for the screening currents in a SQUID with a
sufficiently large self inductance.

\section{Direct detection of the circulating current in a controllable $\pi$-SQUID}

\begin{figure}[t]
\centerline{\psfig{figure=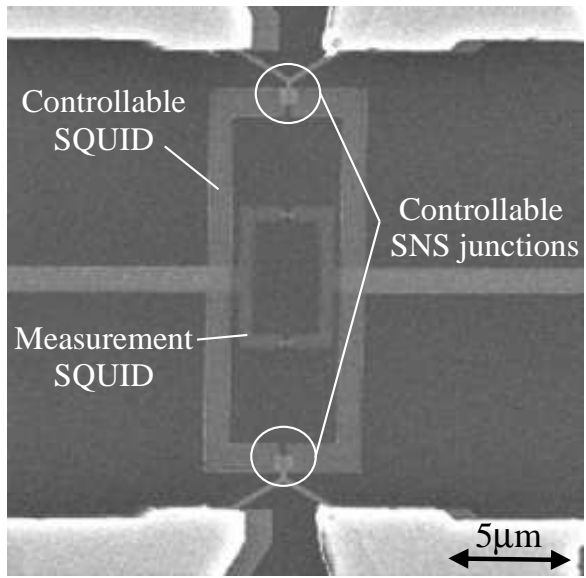,width=8 cm,clip=}}
\caption{Scanning electron microscope of a device that consists of
a controllable $\pi$-SQUID with a smaller conventional DC SQUID
(measurement SQUID) placed inside its loop. The two devices are
electrically isolated by means of 170 nm of Al$_2$O$_3$. The flux
expelled from the big controllable SQUID is detected using the
measurement SQUID.} \label{2squid}
\end{figure}

In this Section we describe an experiment to detect the screening
current in a controllable SQUID directly by placing a conventional
SQUID with small self inductance inside the loop of the
controllable SQUID. The device is shown in Fig.\ref{2squid}. A
controllable SQUID with a large area (70$\mu m^2$) is placed upon
a conventional SQUID, which we call the measurement SQUID, with an
area of 15 $\mu m^2$. This SQUID has conventional SNS junctions
made of gold as weak links. The two devices are electrically
isolated from each other by means of 170 nm of $Al_2O_3$. It is
important to realize that the design of the two SQUIDS is
critical: The product of the screening current J and the self
inductance L of the measurement SQUID should be negligible, to
prevent a disturbance of the controllable SQUID. So, $L\cdot J \ll
\phi_0$. On the other hand the $L\cdot J$ product of the
controllable SQUID should be sufficiently large, so that the flux
expelled is reasonably large. This is particularly important
because the screening current is in the order of or smaller than
the smallest critical current in the system. In a $\pi$-state the
critical current of the junction is smaller than its maximum value
in the normal state (see Fig.\ref{resjie}). If these two
constraints are met we expect a saw tooth behavior of the
circulating current in the controllable SQUID as a function of
$\phi_{ext}$, as shown in Fig.\ref{SQUIDschema}. Hence we expect a
saw tooth behavior of the magnetic field induced in the
measurement SQUID.

The device is made on thermally oxidized silicon. In the first two
steps the measurement SQUID is made using conventional e-beam
lithography with a double layer of PMMA and lift-off. We start
with the normal regions of the junctions, which are made of 40 nm
of gold under which we use 5 nm of titanium as an adhesion layer.
In the second step we DC-sputter deposit the niobium loop (60 nm)
and contacts, which is done after a short \emph{in situ} argon
etching of the gold electrodes. The separation between the Nb
electrode is 450 nm. To isolate the measurement SQUID we cover the
entire structure, except from the contacts (outside the range of
Fig. \ref{2squid}) with 170 nm $Al_2O_3$ using RF sputtering. In
the last two steps we fabricate a controllable SQUID on top of the
small SQUID using the same fabrication procedure as described in
section III.

The experiment is done at 1.4 K. The first step is to determine
the maximum flux generated by the measurement SQUID. To be able to
do this we measure the critical current of the the measurement
SQUID (at $\phi_{ext}=1/2 \phi_0$ and 0) while suppressing the
flux generated by the controllable SQUID. The latter is achieved
by means of applying a control voltage exactly equal to the
critical control voltage to the bottom junction of the
controllable SQUID. This results in a total suppression of the
critical current of this junction, i.e. J=0. Hence we obtain the
critical current of the measurement SQUID at $\phi_{ext}=0$ of 4.2
$\mu A$  and $~$ 200 nA at $\phi_{ext}=0.5 \phi_0$. The maximum
flux generated by the screening current in the measurement SQUID
can be estimated from these measurements using Eq.
\ref{Josephson}, \ref{SQUIDeq} and \ref{fluxeq} to be less than
0.05$\phi_0$, assuming the worst case scenario of exactly
identical critical currents for both junctions \cite{commL}. This
indicates that the measurement SQUID does not influence the
controllable SQUID. In the next step we estimate the self
inductance of the controllable SQUID using the data given in Table
\ref{tab:tablecircle}.

\begin{table}
\caption{\label{tab:tablecircle} Parameters of the controllable
SQUID to estimate the self inductance taken at 1.4 K. $I_{c,MAX}$
= $I_c$ at $\phi_{ext}=0$ and $I_{c,MIN}$=$I_c$ at
$\phi_{ext}=\frac{1}{2} \phi_0$. $I_{c2}$ ($I_{c1}$) is the
critical current of the bottom (top) junction in
Fig.\ref{2squid}.}
\begin{ruledtabular}
\begin{tabular}{llll}
{\bf $I_{c,MAX}$ [$\mu A$]} &{\bf $I_{c,MIN}$ [$\mu A$]} & {\bf
$I_{cB}$ [$\mu A$]} & {\bf $I_{cT}$ [$\mu A$]}
\\ \hline 88 & 76 & 74 & 14 \\
\end{tabular}
\end{ruledtabular}
\end{table}

The critical currents of the bottom and top junctions, $I_{c2}$
and $I_{c1}$ respectively, are measured by applying precisely the
critical control voltage to one junction (reducing its critical
current to 0) and measure the transport critical current through
the SQUID. From these values we can estimate L=65 pH. This value
of L would result in a maximum expelled flux at 1.4 K, if no
control voltages are applied to the junctions, of 0.34 $\phi_0$
associated with $J=11 \mu A$.

\begin{figure}[b]
\centerline{\psfig{figure=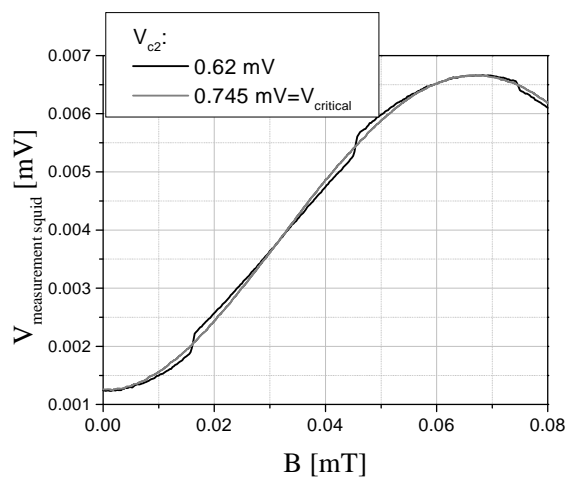,width=8 cm,clip=}}
\caption{Voltage measured over the \emph{measurement} SQUID as a
function of magnetic field for two values of the control voltage
on the bottom junction of the \emph{controllable} SQUID.
$V_{c2}$=0.745 represents the critical control voltage.}
\label{mori}
\end{figure}
\begin{figure}[t]
\centerline{\psfig{figure=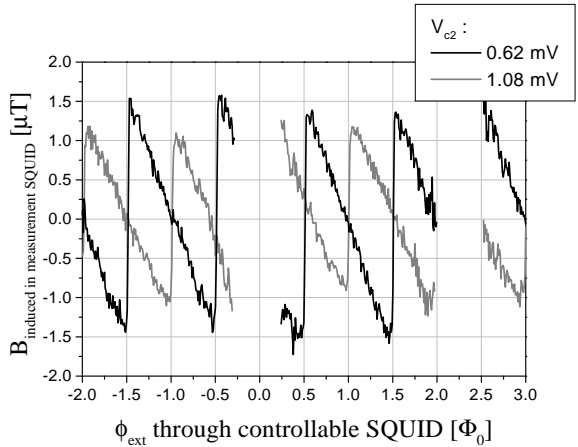,width=8 cm,clip=}}
\caption{The magnetic field induced in the \emph{measurement}
SQUID by the flux generated by the screening current in the
\emph{controllable} SQUID as a function of the external flux given
in units of $\Phi_0$, representing one flux quantum through the
controllable SQUID. The data is given for two different values of
the control voltage over the bottom junction of the
\emph{controllable} SQUID.} \label{sawtooth}
\end{figure}
In the experiment we measure the $V_{SQUID}-B$ oscillations of the
\emph{measurement} SQUID using a suitable low frequency AC bias
current. This is done for different values of $V_{c2}$ on the
\emph{controllable} SQUID. A representative result is shown in
Fig. \ref{mori}. The grey line indicates the situation where
$V_{c2}=V_{c2,critical}=0.745 mV$. In this case $I_{c2}=J$=0. The
result is an (almost) sinuso\"{i}dal voltage with a period of
0.134 mT, which represents $\phi_0$ for the small measurement
SQUID. The black line shows the situation where
$V_{c2}<V_{c2,critical}$=0.62 mV. In this case periodic deviations
from the grey line are observed associated with the screening
current in controllable SQUID. The periodicity of the deviations
is given by $B(\phi_{0,controllable SQUID})\equiv B(\Phi_0)$
=0.029 mT. The ratio
$\frac{B(\Phi_0)}{B(\phi_0)}=\frac{0.029}{0.134}
\approx\frac{15}{70}$ which is the ratio of the areas of the two
SQUIDS. It is this difference in area and hence in periodicity,
that makes it possible to obtain information over the circulating
current in the large controllable SQUID over several fluxquanta
within less than half a fluxquantum in the measurement SQUID. To
make the result more clear we can extract the magnetic field
induced in the measurement SQUID by the screening current in the
controllable SQUID from this data: We subtract the grey curve from
the black one, and multiply this difference by $(\frac{\partial
V}{\partial B})^{-1}$, which is obtained by numerically
differentiating the grey curve. Obviously the data around
$0,\frac{1}{2},1,1\frac{1}{2}$... $\phi_0$ are not obtainable
because the sensitivity of the measurement SQUID is zero at these
points. We perform this analysis on a set of data at different
values of $V_{cB}$. The result is shown in Fig.\ref{sawtooth} for
2 different values of $V_{c2}$ \cite{commVc}. We observe that the
black line, which is the same measurement as shown in Fig.
\ref{mori}($V_{c2}>V_{c2,critical}$=0.62 mV), is a saw tooth
function around integer flux, consistent with the behavior of the
screening current in a SQUID with a relatively large self
inductance, as shown in Fig.\ref{SQUIDschema}c. The grey line, for
which $V_{c2}>V_{c2,critical}$=1.08 mV, shows the exact same
behavior, with however a shift of half a fux quantum. In this case
a screening current flows, in either the clockwise or
counterclockwise direction, at integer external flux.

\begin{figure}
\centerline{\psfig{figure=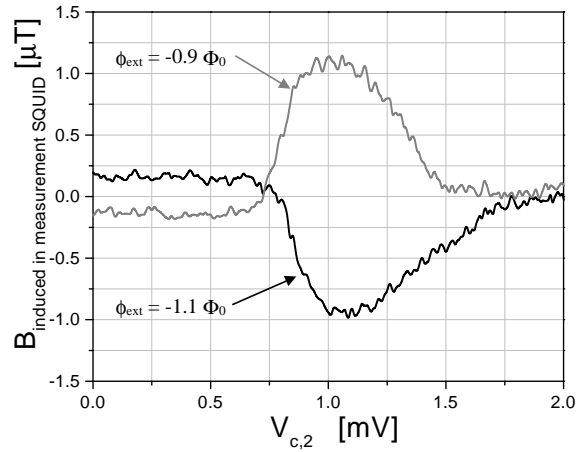,width=8 cm,clip=}}
\caption{The magnetic field induced in the measurement SQUID by
the flux generated by the screening current in the controllable
SQUID as a function of the control voltage applied over the bottom
control channel of the controllable SQUID. The data is shown for 2
values of the external flux $\Phi$ through the \emph{controllable}
SQUID around -1 fluxquantum.} \label{circel}
\end{figure}

These measurements indicate that it is possible, around an integer
flux quantum, to switch between a conventional state without a
screening current and a state with a (bistable) screening current
by applying $V_{c2} > V_{c2,critical}$. This is illustrated more
clearly by the data in Fig.\ref{circel}. In this figure we show a
measurement of the magnetic field induced in the bottom SQUID as a
function of the $V_{c2}$ (in stead of $\phi_{ext}$) for two
different values of the applied external flux around -1
fluxquantum. At $V_{c2}<V_{c2,critical}$ the field induced in the
measurement SQUID is a horizontal line close to 0, indicating that
hardly any flux is expelled from the controllable SQUID. Hence no
screening current flows around the loop of the controllable SQUID
as expected. However, if $V_{c2}>V_{c2,critical}$ we see an
increase (or decrease) of the induced magnetic field, associated
with the screening current caused by the introduction of an extra
phase factor of $\pi$ (-$\pi$) in the SQUID loop. The sign of the
signal depends on the exact value of the external flux. At
$\phi_{ext}=-1.1\phi_0$ (black line), the induced field is
negative, consistent with Fig.\ref{sawtooth}. This clearly
indicates that it is possible to switch from a state with no
current running around the SQUID loop to a non zero current state.
The small a-symmetries in the figure are caused by the magnetic
field associated with the current flowing in the control channel
of the bottom junction.

\section{conclusions}

We have demonstrated that in a controllable $\pi$-SQUID a
screening current can be generated without the application of an
external magnetic field by putting one controllable $\pi$-junction
of the SQUID into a $\pi$-state. At a measurement temperature of
1.4 K and with the present geometry, only a Nb-Ag controllable
$\pi$ junction shows the transition to a $\pi$ state, in contrast
with a Nb-Cu device. We have shown that this difference is caused
by the difference in electron-electron interaction strength
between Ag and Cu, consistent with direct measurements of this
quantity on similar diffusive wires using tunnel junction
spectroscopy \cite{Pothier,Pierre}. The possibility of adding an
extra phase of $\pi$ to the SQUID loop is demonstrated not only by
the screening current, but also by the fact that the $V_{SQUID}-B$
oscillations can be shifted by a phase factor of $\pi$ or 2$\pi$,
depending wether one or two controllable junctions are put on a
$\pi$-state.

We gratefully acknowledge H. Pothier and F.K. Wilhelm for
discussions and for making their computer programs available to
us. This work was supported by the Nederlandse Organisatie voor
Wetenschappelijk Onderzoek (NWO) through the Stichting voor
Fundamenteel Onderzoek der Materie (FOM).


\begin{thebibliography}{99}
\bibitem{critcur} We have used the fact that the
critical current generally  equals the supercurrent if the phase
difference is $\pi$/2. This is not fully correct, but only leads
to minor changes in the calculated curves \cite{Wilhelm,Yip}.
\bibitem{commheating} The dominant thermal resistance in our system is the one
associated with electron heat diffusion from the control channel
into the reservoirs \cite{Henny}. We estimated the electron
diffusion heating parameter $b$ valid for our samples at 1.4 K,
$b\leq0.05 k/mV$. The calculated effective electron temperature in
the reservoirs $T_{eff}=\sqrt{{T_{bath}^2}+{bV_c}^2}$, yields a
negligible electron heating in the range of $V_c$ used in the
experiments $0 \leq V_c \leq 1.2$ mV.
\bibitem{commGe} The use of a Ge suspended mask,
which is more stable than the presented PMMA mask, is not suitable
because the electron beam exposure of the large area reservoirs
has to be done using relatively high beam currents ($~$nA), to
come to a reasonable exposure time. This results in stress in the
PMMA-MA-Ge-PMMA resist, causing the germanium to crack.
\bibitem{commcu} We calculated $I_c$ as a function of $\tau_D$
with identical parameters and found that at $\tau_D$=2ns the
trasnition to a $\pi$-state would occur at 1.4K. At $T<1 K$ the
transition would even occur with the sample as it is.
\bibitem{commL} The self inductance would be 45 pH in the case of
identical junctions. However, the difference in transport critical
current between $\phi=0$ and $\phi=\phi_0$/2 can also be
attributed to small difference in critical currents between the
junctions. This assumption gives L=0 and a difference in critical
current of 10 percent. We must conclude that L$\leq$45pH for the
measurement SQUID.
\bibitem{commVc} In Fig. \ref{sawtooth} we have corrected for a
slight horizontal and vertical shift between the curves that is
caused by the current associated with $V_{c2}$ flowing through the
control channel. The flux associated with this current causes a
horizontal shift between the curves due to the coupling in the
controllable SQUID, identical to the one in Fig.\ref{oscillations}
and a vertical shift due to the coupling into the measurement
SQUID.
\end{thebibliography}
\end{document}